\documentstyle{article}

\newtheorem{theorem}{Theorem}[section]
\newtheorem{lemma}[theorem]{Lemma}

\newtheorem{definition}[theorem]{Definition}
\newtheorem{proposition}[theorem]{Proposition}

\newtheorem{remark}[theorem]{Remark}

\def\t{\hbox}
\def\il{\int\limits}  
\def\g{\gamma}
\def\f{\frac} 
\def\q{\quad}
\def\p{\varphi}  

\def\P{\Phi}

\def\a{\alpha}
\def\b{\beta} 
\def\s{\sigma}
\def\ep{\varepsilon}
   
\def\ba{\begin{array}}
\def\pr{\prime}
\def\y{\tilde{y}}

\begin{document}
\begin{center}

{\bf Continuous methods for solving nonlinear ill-posed problems}\\
\vspace{0.5cm}

Ruben G. Airapetyan\\
E-mail: airapet@math.ksu.edu\\
Department of Mathematics\\
Kansas State University\\Manhattan, Kansas 66506-2602, U.S.A.\\
\vspace{0.3cm}

Alexander G. Ramm\\
E-mail: ramm@math.ksu.edu\\
Department of Mathematics\\
Kansas State University\\Manhattan, Kansas 66506-2602, U.S.A.\\
\vspace{0.3cm}

Alexandra B. Smirnova\\
E-mail: smirn@math.ksu.edu\\
Department of Mathematics\\
Kansas State University\\Manhattan, Kansas 66506-2602, U.S.A.\\
\end{center}

\begin{abstract}
The goal of this paper is to develop a general approach
to solution of ill-posed nonlinear problems in a Hilbert
space based on continuous processes with a regularization
procedure. To avoid the ill-posed inversion of the Fr\'echet
derivative operator a regularizing one-parametric family of operators is
introduced. Under certain assumptions on the regularizing family a general
convergence theorem is proved. The proof is based on a lemma describing
asymptotic behavior of solutions of a new nonlinear integral 
inequality. Then the applicability of the theorem to the continuous analogs
of the Newton, Gauss-Newton and simple iteration methods is demonstrated.  


\end{abstract}

\section{Introduction}
\setcounter{equation}{0}
\setcounter{theorem}{0}
\renewcommand{\thetheorem}{1.\arabic{theorem}}
\renewcommand{\theequation}{1.\arabic{equation}}
\vspace*{-0.5pt}

Let us consider a nonlinear operator equation
\begin{equation}\label{noneq}
F(z)=0,\q F: H\to H, 
\end{equation}
in a real Hilbert space $H$ (equation (\ref{noneq}) 
in a complex Hilbert space can
be treated similarly).

Assume that (\ref{noneq}) is solvable (not necessarily uniquely). 
If the Fr\'echet derivative of the operator $F$ has nontrivial null-space
at the solution to (\ref{noneq}), then one can use the classical Newton 
method for solution to (\ref{noneq}) only under some strong assumptions on the
operator $F$ (see \cite{dkk,ag}). Otherwise in order to construct a numerical
method for solution to (\ref{noneq}) one needs some regularization
procedure. 

In the theory of ill-posed problems many different discrete methods based on a
regularization are known. Many different convergence theorems for such
schemes describe the efficiency of the numerical algorithms for solving
various nonlinear problems, and give existence results (see, for example,
\cite{bns,ehn,va}). However it is
quite difficult to navigate in the sea of the discrete schemes and
corresponding convergence theorems. Proofs of these theorems are usually
based on the contraction mapping principle and are sometimes rather   
complicated. 

On the other hand an analysis of continuous processes is  
based on the investigation of the asymptotical behavior of 
nonlinear dynamical systems in Banach and Hilbert spaces. If a  
convergence theorem is proved for a continuous method, 
one can construct various 
discrete schemes generated by this continuous process. Thus
construction of a discrete numerical scheme is split into two parts:   
construction of the continuous process and numerical integration of the
corresponding nonlinear operator differential equation. Convergence theorems for
regularized continuous Newton-like methods are established in \cite{ars,al,r}. 

The goal of this paper is to develop a general approach to continuous analogs of
discrete methods and to establish fairly general convergence theorem.  This approach
is based on an analysis of the solution to the Cauchy
problem for a nonlinear differential equation in a Hilbert space. Such an analysis
was done for well-posed problems in \cite{a}, where it was based on a usage of an
integral inequality.  It is more difficult to study nonlinear ill-posed problems.  In
this case one has to use more complicated new integral inequality (Lemma~\ref{ric}).
Based on Lemma~\ref{ric} the general theorem establishing convergence of a
regularized continuous process is proved (Theorem~\ref{t1}). Applying this theorem to
the regularized Newton's and simple iteration methods (for monotone operators) and to
Gauss-Newton-type methods (for non-monotone operators) convergence theorems are
obtained under less restrictive conditions on the equation than the theorems known
for the corresponding discrete methods. According to these theorems
one can choose a regularizing operator depending on the "degree of
degeneracy" of the original nonlinear problem and estimate the rate of
convergence of the regularized process.

The paper is organized as follows. In Section 2 Lemma~\ref{ric} about the solution of
a new integral inequality and Theorem~\ref{t1} about the convergence of a regularized
continuous process are formulated and proved. In Section 3 this theorem is applied to
continuous Newton's and simple iteration methods and in Section 4 to the
Gauss-Newton-type methods. A practically interesting numerical example is considered
in Section 5. A lemma about nonlinear differential inequality is proved              
in the Appendix.

\section{Regularization procedure}
\setcounter{equation}{0}
\setcounter{theorem}{0}
\renewcommand{\thetheorem}{2.\arabic{theorem}}
\renewcommand{\theequation}{2.\arabic{equation}}
\vspace*{-0.5pt}

In the well-posed case (the Fr\'echet derivative $F'$ of the operator $F$ is a
bijection in a neighborhood of the solution of equation (\ref{noneq})) in order
to solve equation (\ref{noneq}) one can use the following continuous processes:
\begin{itemize}
\item simple iteration method:
\begin{equation}\label{esim2}
\dot{z}(t)=-F(z(t)),\q z(0)=z_0\in H,
\end{equation}
\item Newton's method:
\begin{equation}\label{enm2}
\dot{z}(t)=-[F'(z(t))]^{-1}F(z(t)),\q z(0)=z_0\in H,
\end{equation}
\item Gauss-Newton's method:
\begin{equation}\label{egnm}
\dot{z}(t)=-[F^{\pr *}(z(t))F^{\pr}(z(t))]^{-1}F^{\pr *}(z(t))F(z(t)),\q
z(0)=z_0\in H,
\end{equation}
\end{itemize}
or some of their modifications (see \cite{a,g}).
  
However if $F'$ is not continuously invertible (ill-posed case) one has to
replace equations (\ref{esim2}) -(\ref{egnm}) by the corresponding 
regularized equations:
\begin{itemize}
\item regularized simple iteration method:
\begin{equation}\label{esim1}
\dot{z}(t)=-[F(z(t))+\ep(t)(z(t)-z_0)],\q z(0)=z_0\in H,
\end{equation} 
\item regularized Newton's method:
\begin{equation}\label{enm1} 
\dot{z}(t)=-[F'(z(t))+\ep(t)I]^{-1}[F(z(t))+\ep(t)(z(t)-z_0)],\q z(0)=z_0\in H,
\end{equation}
\item regularized Gauss-Newton's methods: equation (\ref{caupr}) with the function
$\P$ defined in (\ref{P1}) or in (\ref{P2}),
\end{itemize}
with an appropriate choice of the function $\ep(t)$ and the point $z_0$. Here $I$ is
the identity operator.

The goal of this paper is to develop a uniform approach to such regularized
methods. Let us consider the Cauchy problem:
\begin{equation}\label{caupr}
\dot{z}(t)=\P(z(t),t),\q z(0)=z_0\in H,
\end{equation}
with an operator $\P:H\times[0,\infty)\to H$. The choice of $\P$ yields
the corresponding continuous process.

In this section a general convergence theorem (Theorem~\ref{t1}) is established. 
In the next two
sections the convergence theorems for the processes mentioned above are derived from  
this general theorem. In the proof of the general theorem the technique of integral
inequalities is used.

The following lemma is known. It is a version 
of some results concerning integral
inequalities (see e.g. Theorem 22.1 in \cite{JS}). 
For convenience of the reader and to make the presentation essentially
self-contained we include a proof.

\begin{lemma}  
Let $f(t,w)$, $g(t,u)$ be continuous on region $[0,T)\times D$ ($D\subset
R$, $T\le\infty$) and $f(t,w)\leq g(t,u)$ if $w\leq u$,  $t\in (0,T)$, $w,u\in D$.
Assume that 
$g(t,u)$
is such that the Cauchy problem 
\begin{equation}\label{Cauchy}
\dot u = g(t,u), \q u(0)=u_0,\q u_0\in D
\end{equation}
has a unique solution. If
\begin{equation}\label{inequal}
\dot w\leq f(t,w), \q w(0)=w_0\leq u_0,\q w_0\in D,
\end{equation}
then $u(t)\geq w(t)$ for all $t$ for which $u(t)$ and $w(t)$ are
defined.
\label{ineq}\end{lemma}

{\bf Proof} \underline{Step 1.} Suppose first $f(t,w)<g(t,u)$, if $w\leq u$.
Since 
$w_0\leq u_0$ and $\dot w (0)\leq f(t,w_0)<g(t,u_0)=\dot u (0)$, there exists $\delta
>0$ such that $u(t)> w(t)$ on $(0,\delta]$. Assume that for some $t_1> \delta$ one
has
$u(t_1)< w(t_1).$ Then for some $ t_2<t_1$ one has
$$
u(t_2)=w(t_2) \q\t{and}\q
u(t)<w(t) \q\t{for}\q  t\in (t_2,t_1].
$$
One gets
$$
\dot w(t_2)\geq \dot u(t_2)=g(t,u(t_2))>f(t,w(t_2))\geq \dot w(t_2).
$$
This contradiction proves that there is no point $t_2$ such that $u(t_2)=w(t_2)$.
 
\underline{Step 2.} Now consider the case $f(t,w)\leq g(t,u)$, if $w\leq u$.  
Define 
$$
\dot u_n = g(t,u_n)+\ep_n, \q u_n(0)=u_0,\q \ep_n>0,\q n=0,1,...,
$$
where $\ep_n$ tends monotonically to zero. Then
$$
\dot w\leq f(t,w)\leq g(t,u)<g(t,u)+\ep_n, \q w\leq u.
$$
By Step 1 $u_n(t)\geq w(t)$, $n=0,1,...\, $. Fix an arbitrary compact set
$[0,T_1]$, $0<T_1<T.$ 
\begin{equation}\label{un}
u_n(t)=u_0+\il_0^tg(\tau,u_n(\tau))d\tau + \ep_n t.
\end{equation}
Since $g(t,u)$ is continuous, the sequence $\{u_n\}$ is uniformly
bounded and
equicontinuous on $[0,T_1]$. Therefore there exists a subsequence $\{u_{n_k}\}$
which converges uniformly to a continuous function $u(t)$. By continuity of $g(t,u)$
we
can pass to
the limit in  (\ref{un}) and get 
\begin{equation}\label{un1a}
u(t)=u_0+\il_0^tg(\tau,u(\tau))d\tau,\q t\in [0,T_1].
\end{equation}
Since $T_1$ is arbitrary (\ref{un1a}) is equivalent to the initial Cauchy problem
that has a unique solution. 
The inequality $u_{n_k}(t)\geq w(t),$ $k=0,1,...$ implies $u(t)\geq w(t).$
If the solution to the Cauchy problem (\ref{Cauchy}) is not unique, the inequality
$w(t)\leq u(t)$ holds for the maximal solution to (\ref{Cauchy}). 
$\Box$

Our second lemma is a key to the basic result of this section, namely to 
Theorem~\ref{t1}. 

\begin{lemma} 
Let $\g(t), \s(t), \b(t)\in C[0,\infty)$.
If there exists a positive  function $\mu(t)\in C^1[0,\infty)$ such that
\begin{equation}\label{ric1}
0\le\s(t)\le\f{\mu(t)}{2}\left(\g(t)-\f{\dot{\mu}(t)}{\mu(t)}\right),\q
\b(t)\le\f{1}{2\mu(t)}\left(\g(t)-\f{\dot{\mu}(t)}{\mu(t)}\right),\q
\mu(0)v(0)<1,
\end{equation}
then a nonnegative solution to the following inequality
\begin{equation}\label{ric4}
\dot{v}(t)\le -\g(t) v(t)+\s(t)v^2(t)+\b(t).
\end{equation}
satisfies the estimate:
\begin{equation}\label{ric5}
v(t)\,<\, \f{1}{\mu(t)}.
\end{equation}
\label{ric}\end{lemma}

\begin{remark}
Without loss of generality one can assume $\b(t)\geq 0$. 

In \cite{al1} (see also Appendix) a differential inequality
$\dot{u}\le -a(t)\psi(u(t))+b(t)$ was studied under assumptions
1) - 3) of Lemma~\ref{apineq} of the Appendix. These assumptions alone,
as we show in the Appendix, do not imply the desired conclusion
(\ref{descon}). We have added assumption 4) in order to prove conclusion
(\ref{descon}).  In Lemma~\ref{ric}
the term $-\g(t)v(t)+\s(t)v^2(t)$ (which is analogous to some extent to the term
$-a(t)\psi(u(t))$) can change sign. Our Lemma~\ref{ric} is not covered by the
result in \cite{al1}. In particular, in Lemma~\ref{ric} an analog of $\psi(u)$,
for the case $\g(t)=\s(t)=a(t)$, is the function $\psi(u):=u-u^2$.
This function goes to $-\infty$
as $u$ goes to $+\infty$, so it does not satisfy the positivity
condition imposed in \cite{al1}.

Unlike in the case
of Bihari integral inequality (\cite{bb}) one cannot separate variables 
in the right
hand side of inequality (\ref{ric4}) and estimate $v(t)$
by a solution of the Cauchy problem for a differential equation with separating
variables. The proof below is based on a special choice of the solution to
the Riccati equation majorizing a solution of integral inequality (\ref{ric4}).
\label{bihari}\end{remark}

{\bf Proof}{\bf of Lemma~\ref{ric}} Denote:
\begin{equation}\label{rc1}
w(t):=v(t)e^{\int_0^t \g(s)ds},
\end{equation}
then (\ref{ric4}) implies:
\begin{equation}\label{rc2}
\dot{w}(t)\le a(t)w^2(t)+b(t),\q w(0)=v(0),
\end{equation}
where
$$ 
a(t)=\s(t)e^{-\int_0^t \g(s)ds},\q b(t)=\b(t)e^{\int_0^t \g(s)ds}.
$$
Consider Riccati's equation:
\begin{equation}\label{rc3}
\dot{u}(t)=\f{\dot{f}(t)}{g(t)}u^2(t)-\f{\dot{g}(t)}{f(t)}.
\end{equation}
One can check by a direct calculation that the the solution to
problem~(\ref{rc3}) is given by the following formula 
\cite[eq. 1.33]{kam}:
\begin{equation}\label{rc5}
u(t)=-\f{g(t)}{f(t)}+\left[f^2(t)\left(C-
\int_0^t\f{\dot{f}(s)}{g(s)f^2(s)}ds\right)\right]^{-1}.
\end{equation}
Define $f$ and $g$ as follows:
\begin{equation}\label{rc4}
f(t):=\mu^{\f{1}{2}}(t)e^{-\f{1}{2}\int_0^t \g(s)ds},\q
g(t):=-\mu^{-\f{1}{2}}(t)e^{\f{1}{2}\int_0^t \g(s)ds},
\end{equation}
and consider the Cauchy problem for equation 
(\ref{rc3}) with the initial condition
$u(0)=v(0)$. Then $C$ in (\ref{rc5}) takes the form:
$$C=\f{1}{\mu(0)v(0)-1}.$$
From (\ref{ric1}) one gets
$$
a(t)\le \f{\dot{f}(t)}{g(t)},\q b(t)\le -\f{\dot{g}(t)}{f(t)}.
$$ 
Since $fg=-1$ one has:
$$
\int_0^t\f{\dot{f}(s)}{g(s)f^2(s)}ds=-\int_0^t\f{\dot{f}(s)}{f(s)}ds=
\f{1}{2}\int_0^t\left(\g(s)-\f{\dot{\mu}(s)}{\mu(s)}\right)ds.
$$   
Thus
\begin{equation}\label{rc6}
u(t)=\f{e^{\int_0^t \g(s)ds}}{\mu(t)}\left[1-\left(\f{1}{1-\mu(0)v(0)}+
\f{1}{2}\int_0^t\left(\g(s)-\f{\dot{\mu}(s)}{\mu(s)}\right)ds\right)^{-1}\right].
\end{equation}
It follows from condition (\ref{ric1}) that the solution to
problem
(\ref{rc3}) exists for all $t\in[0,\infty)$ and the following inequality holds:
\begin{equation}\label{estimate}
1\, >\, 1-\left(\f{1}{1-\mu(0)v(0)}+
\f{1}{2}\int_0^t\left(\g(s)-\f{\dot{\mu}(s)}{\mu(s)}\right)ds\right)^{-1}
\ge\mu(0)v(0).
\end{equation}
From Lemma~\ref{ineq} and from formula (\ref{rc6}) one gets:
\begin{equation}\label{rc7}
v(t)e^{\int_0^t \g(s)ds}:=
w(t)\le u(t)<\f{1}{\mu(t)}e^{\int_0^t \g(s)ds},
\end{equation}
and thus estimate (\ref{ric5}) is proved.
$\Box$

{\bf Examples.} To illustrate conditions (\ref{ric1}) of Lemma~\ref{ric} consider the
following examples of functions $\g$, $\s$, $\b$, satisfying (\ref{ric1}). 

{\bf 1.} Let
\begin{equation}\label{rcex} 
\g(t)=c_1(1+t)^{\nu_1},\q \s(t)=c_2(1+t)^{\nu_2},\q \b(t)=c_3(1+t)^{\nu_3},
\end{equation}
where $c_2>0$, $c_3>0$. 
Choose $\mu(t):=c(1+t)^{\nu}$, $c>0$. From (\ref{ric1}) one gets the following
conditions
$$
c_2\le \f{cc_1}{2}(1+t)^{\nu+\nu_1-\nu_2}-\f{c\nu}{2}(1+t)^{\nu-1-\nu_2},\q
$$
\begin{equation}\label{rcex1}
c_3\le \f{c_1}{2c}(1+t)^{\nu_1-\nu-\nu_3}-\f{\nu}{2c}(1+t)^{-\nu-1-\nu_3},\q
cv(0)<1.  
\end{equation}
Thus one obtains the following conditions:
\begin{equation}\label{rcex2}
\nu_1\ge -1,\q \nu_2-\nu_1\le\nu\le\nu_1-\nu_3,
\end{equation}
and
\begin{equation}\label{rcex3}
c_1>\nu,\q \f{2c_2}{c_1-\nu}\le c\le\f{c_1-\nu}{2c_3},\q
cv(0)<1.
\end{equation}
Therefore for such $\g$, $\s$, $\b$ a function $\mu$ with the desired 
properties exists if
\begin{equation}\label{rcex4}
\nu_1\ge -1,\q \nu_2+\nu_3\le 2\nu_1,  
\end{equation}
and
\begin{equation}\label{rcex5}
c_1>\nu_2-\nu_1,\q 2\sqrt{c_2c_3}\le c_1+\nu_1-\nu_2,\q
2c_2v(0)<c_1+\nu_1-\nu_2.
\end{equation}
In this case one can choose $\nu=\nu_2-\nu_1$, $c=\f{2c_2}{c_1+\nu_1-\nu_2}$.
However in order to have $v(t)\to 0$ as $t\to+\infty$ (the case of interest
in Theorem~\ref{t1}) one needs the following conditions:
\begin{equation}\label{rcex6}
\nu_1\ge -1,\q \nu_2+\nu_3\le 2\nu_1,\q \nu_1>\nu_3,
\end{equation}
and
\begin{equation}\label{rcex7}
c_1>\nu_2-\nu_1,\q 2\sqrt{c_2c_3}\le c_1,\q 2c_2v(0)<c_1.
\end{equation}
 
{\bf 2.} If 
$$
\g(t)=\g_0,\q \s(t)=\s_0e^{\nu t},\q \b(t)=\b_0e^{-\nu t},\q \mu(t)=\mu_0e^{\nu t},
$$
then conditions (\ref{ric1}) are satisfied if 
$$
0\le \s_0\le\f{\mu_0}{2}(\g_0-\nu),\q\b_0\le\f{1}{2\mu_0}(\g_0-\nu),\q\mu_0v(0)<1.
$$

{\bf 3.} If 
$$
\g(t)=\f{1}{\sqrt{\log(t+t_0)}},\q \mu(t)=c\log(t+t_0),
$$
then conditions (\ref{ric1}) are satisfied if
$$  
0\le\s(t)\le\f{c}{2}\left(\sqrt{\log(t+t_0)}-\f{1}{t+t_0}\right),
$$
$$
\b(t)\le\f{1}{2c\log^2(t+t_0)}\left(\sqrt{\log(t+t_0)}-\f{1}{t+t_0}\right),\q
v(0)c\log t_0<1.
$$

In all considered examples $\mu(t)$ can
tend to infinity as $t\to +\infty$ and  provide a decay of a nonnegative
solution to integral inequality (\ref{ric4}) even if $\s(t)$ tends to infinity.
Moreover in the first and the third examples $v(t)$ tends to zero as $t\to
+\infty$ when $\g(t)\to 0$ and $\s(t)\to +\infty$.
 
\begin{theorem}   
Let $\P(h,t)$ be Fr\'echet differentiable with respect to $h$ and satisfy the
following condition:

there exists a differentiable function $x(t)$, $x:[0,+\infty)\to H$, such that
for any $h\in H,\, t\in[0,+\infty)$
\begin{equation}\label{semon}
(\P(h,t),h-x(t))\le \a(t)||h-x(t)||-\g(t)||h-x(t)||^2+\s(t)||h-x(t)||^3, 
\end{equation}
where $\alpha(t)$ is a continuous function, $\alpha(t)\ge 0$, $\g(t)$ and $\s(t)$
satisfy conditions (\ref{ric1}) of Lemma~\ref{ric} with 
\begin{equation}\label{xdot}
\b(t):=||\dot{x}(t)||+\a(t),\q v(0):=||z_0-x(0)||,\q v(t):=||z(t)-x(t)||,
\end{equation}
and $\mu(t)$ tends to $+\infty$ as $t\to +\infty$.

Then problem (\ref{caupr}) has a unique solution $z(t)$ defined for all $t\in
[0,\infty)$, and
\begin{equation}\label{solut}
||z(t)-x(t)||<\frac 1 {\mu(t)},\q \lim_{t\to+\infty}||z(t)-x(t)||=0.
\end{equation}
\label{t1}\end{theorem}

\begin{remark}
One can choose the regularizing operator $\P(h,t)$ in (\ref{caupr})
such that condition (\ref{semon}) holds in the case when 
$F^{\pr *}(h)F'(h)$ is not boundedly invertible (see Sect. 3).   
\label{nobound}\end{remark}

{\bf Proof} {\bf of Theorem~\ref{t1}} Since $\P(h,t)$ is Fr\'echet differentiable
with respect to $h$
there exists the solution to problem (\ref{caupr}) on the maximal interval $[0,T_1)$
of the existence of the solution to (\ref{caupr}).
One has to show that $T_1=+\infty$. Assume $T_1<+\infty$. 
Since $H$ is a real Hilbert space one has:
\begin{equation}\label{eneq}
\f{1}{2}\f{d}{dt}||z(t)-x(t)||^2=(\dot{z}-\dot{x},z(t)-x(t))
=(\P(z(t),t),z(t)-x(t))-(\dot{x},z(t)-x(t)).
\end{equation}
Therefore from (\ref{semon}) and (\ref{xdot}) one obtains
\begin{equation}\label{est1}
\f{1}{2}\f{d}{dt}||z(t)-x(t)||^2\le
-\g||z(t)-x(t)||^2+\s(t)||z(t)-x(t)||^3+\b(t)||z(t)-x(t)||.
\end{equation}
Denote
$$
v(t):=||z(t)-x(t)||.
$$
From (\ref{est1}) one has:
$$
v(t)\dot{v}(t)\le -\g(t)v^2(t)+\s(t)v^3(t)+\b(t)v(t). 
$$
If $v>0$, one gets:
\begin{equation}\label{ineq11}
\dot{v}(t)\le -\g(t)v(t)+\s(t)v^2(t)+\b(t).
\end{equation}
If $v=0$ on some interval, then inequality (\ref{ineq11})
is satisfied trivially because $\beta(t)\ge 0$.
Thus (\ref{ineq11}) holds for all $t>0$.

By Lemma~\ref{ric} one obtains
\begin{equation}\label{appr}
||z(t)-x(t)||\le\f{1}{\mu(t)},\q \t{ for } \q t\in[0,T_1).
\end{equation}
From (\ref{appr}) one concludes that $z(t)$ does not leave the ball
$B_1$ centered at $x(t)$
  with radius $(\min_{t\in [0,T_1]}\mu(t))^{-1}$ $ > 0$.
Since $\max_{0\leq t \leq T_1}||x(t)||<\infty$,  one
concludes that $\sup_{0\leq t < T_1}||z(t)||<\infty$.
Therefore there exists a sequence $\{t_n\}\to T_1$ such
 that $\{z(t_n)\}$ converges
weakly
to some $\tilde{z}$. From equation (\ref{caupr}) one 
derives the uniform boundedness of the norm $||\dot{z}(t)||$ 
on $[0,T_1)$. Thus there exists 
$\lim_{t\to T_1}||z(t)-\tilde{z}||=0$. Since
the
conditions for the uniqueness and
local solvability of the Cauchy problem for equation (\ref{caupr})
with initial condition $z(T_1)=\tilde{z}$ are satisfied, one can continue
the solution to (\ref{caupr}) through $T_1$. This contradicts the assumption of
maximality of $T_1$, thus $T_1=+\infty$.
Moreover, from (\ref{ric5}) one gets:   
\begin{equation}\label{est2}
\lim_{t\to+\infty}||z(t)-x(t)||\le \lim_{t\to+\infty}\f{1}{\mu(t)}=0.
\end{equation}
$\Box$

\section{Regularized Continuous Methods\\ for Monotone Operators}
\setcounter{equation}{0}
\setcounter{theorem}{0}
\renewcommand{\thetheorem}{3.\arabic{theorem}}
\renewcommand{\theequation}{3.\arabic{equation}}
\vspace*{-0.5pt}

In this section we apply the regularization procedure described in Sect. 2
to solve nonlinear operator equation (\ref{noneq}). Assume that $F$ is
Fr\'echet differentiable and
\begin{equation}\label{frder}
(F'(h)\xi,\xi)\ge 0\q\t{ for all }\q h,\xi\in H.
\end{equation}
Under this assumption the operator $F'(h)+\ep(t)I$ is boundedly
invertible.
Define $\P$ as follows:
\begin{equation}\label{rchn}
\P(h,t):=-[F'(h)+\ep(t)I]^{-1}[F(h)+\ep(t)(h-z_0)],
\end{equation}
where $z_0\in H$ is an initial approximation point and
$\ep(t)$ is some positive function on the interval $[0,\infty)$. Some restrictions on
$\ep(t)$ will be stated in Theorem~\ref{t3}.

An outline of the convergence proof is the following. First one considers an
auxiliary well-posed problem:
\begin{equation}\label{pr1}
F_\ep(x):=F(x)+\ep(x-z_0)=0, \q \ep>0,
\end{equation}
and shows that the difference between its solution $x(t)$ and the solution $z(t)$ to problem
(\ref{caupr}) tends to zero as $t\to +\infty$. On the other hand one shows that
$x(t)$ converges to the exact solution $y$ of equation (\ref{noneq}). Thus one proves
the convergence of $z(t)$ to $y$ as $t\to +\infty$.

We recall first some definitions from nonlinear functional analysis which are used below.
The most essential restrictions on the operator $F$ imposed in this section are
(\ref{frder}) and {\it w}-continuity of $F$. In particular they imply monotonicity      
and hemicontinuity of $F$.
 
\begin{definition}  
A mapping $\p$ is monotone in a Hilbert space $H$ if
$$
(\p(x_1)-\p(x_2),x_1-x_2)\ge 0,\q \forall x_1,x_2\in H.
$$
\end{definition}
\begin{definition}
A mapping $\p$ is hemicontinuous at $x_0\in H$ if the
map $t\to (\p(x_0+th_1),h_2)$ is continuous in a neighborhood of $t=0$
for any
$h_1,h_1\in H$.   
\end{definition}
\begin{definition}
A mapping $\p$ is strongly monotone in a Hilbert space $H$ if
there exists a constant $k>0$ such that
$$
(\p(x_1)-\p(x_2),x_1-x_2)\ge k||x_1-x_2||^2,\q \forall x_1,x_2\in H.
$$
\end{definition}
\begin{lemma}
If $F$ is monotone and hemicontinuous then the problem (\ref{pr1})
is uniquely solvable.
\label{l1}\end{lemma}

{\bf Proof} {\bf of Lemma~\ref{l1}}
According to \cite[p. 100]{deim} problem (\ref{pr1}) is solvable if the operator
$F_\ep$ is  monotone, hemicontinuous and $||F_\ep(x)||\to\infty$ as $||x||\to\infty$.
For sufficiently small $\delta>0$ from  (\ref{pr1}) one has:
$$
||F_\ep(x)||^2\ge
||F(x)||^2+\ep^2||x-z_0||^2-\f{1}{4\delta}||F(z_0)||^2-\delta||x-z_0||^2
$$
$$
\ge(\ep^2-\delta)||x-z_0||^2-C\to +\infty
$$
as $||x||\to+\infty$. Here $C$ is a constant.
Since $F_\ep$ is strongly monotone the solution to (\ref{pr1}) is unique.
Therefore Lemma ~\ref{l1} is proved.
$\Box$

\begin{remark}
The result given by Lemma~\ref{l1} is well known and
its proof is given for the convenience of the reader.
\end{remark}

Let $\rightharpoonup$ denote weak convergence in $H$.
  
\begin{definition}
We say that $F$ is {\it w}-continuous if $x\rightharpoonup \xi$ implies $F(x)
\rightharpoonup F(\xi)$.
\end{definition}

\begin{lemma}
Suppose  that   $F$ is  w-continuous, all the assumptions of Lemma  ~\ref{l1}
are satisfied, and 
there exists a unique solution $y$ to
(\ref{noneq}).
Let  $x(t)$ solve (\ref{pr1}) for $\ep=\ep(t)$, and $\ep(t)$
tend to zero as $t\to +\infty$. Then
\begin{equation}\label{znorms}
\lim_{t\to+\infty}||x(t)-y||=0.
\end{equation}
\label{l2}\end {lemma}

{\bf Proof}
First let us show that $x(t)$ is bounded. Indeed, it follows from (\ref{pr1}) 
that   
$$
F(x(t))-F(y)+\ep(t)(x(t)-y)=\ep(t)(z_0-y).
$$
Therefore
\begin{equation}\label{3.6}
(F(x(t))-F(y), x(t)-y)+\ep(t)||x(t)-y||^2=\ep(t)(z_0-y,x(t)-y).
\end{equation}
This and (\ref{frder}) imply
\begin{equation}\label{toobad}
||x(t)-y||\le||z_0-y||.
\end{equation}
Thus there exists a sequence $\{x(t_n)\}$, $t_n\to\infty$ as $n\to\infty$,
which  converges weakly to some element $\y\in H$. Let us show that $\y$ is the
(unique) solution to problem (\ref{noneq}). Since $F$ is {\it w}-continuous,
$F(x(t_n))\rightharpoonup F(\y)$. Because of the weak lower semicontinuity 
of the norm in a Hilbert space one has:
\begin{equation}\label{wsol}
||F(\y)||\le\liminf_{n\to\infty}||F(x(t_n))||=\liminf_{n\to\infty}\ep(t_n)||x(t_n)-z_0||
=0.
\end{equation}
The conclusion $F(\y)=0$ follows from (\ref{wsol}) and can also be derived directly
from (\ref{pr1}) with $\ep=\ep(t)\to 0$ as $t\to +\infty$. By the uniqueness of the
solution to equation (\ref{noneq}) one concludes that 
$\y=y$. Let us show that the sequence $\{x(t_n)\}$ converges strongly to $y$.
Indeed, from (\ref{3.6}), (\ref{frder}) and the relation $x(t_n)\rightharpoonup y$,
one gets:
\begin{equation}\label{ineq1}
||x(t_n)-y||^2\le (z_0-y,x(t_n)-y)\to 0\t{ as }n\to\infty.
\end{equation}
Thus
\begin{equation}\label{lim1}   
\lim_{n\to\infty}||x(t_n)-y||=0.
\end{equation}
From (\ref{lim1}) it follows by the standard argument that $x(t)\to y$
as
$t\to\infty$. Lemma~\ref{l2} is
proved.
$\Box$

\begin{lemma}
Assume  that $F$ is continuously Fr\'echet differentiable, 
$\sup_{x\in H}||F'(x)||\le N_1$,
and condition (\ref{frder}) holds. If $\ep(t)$ is continuously differentiable,
then the solution $x(t)$ to problem (\ref{pr1}) with $\ep=\ep(t)$ is continuously
differentiable in
the strong sense and one has
\begin{equation}\label{est8}
||\dot{x}(t)||\le\f{|\dot{\ep}(t)|}{\ep(t)}||y-z_0||,\q t\in [0,+\infty).
\end{equation}
\label{l3}\end {lemma}

{\bf Proof}
 Fr\'echet differentiability of $F$ implies hemicontinuity of $F$. Therefore 
problem (\ref{pr1}) with $\ep=\ep(t)$ is uniquely solvable.  
The differentiability of $x(t)$ with respect to $t$ 
follows from the implicit function
theorem [2]. To derive (\ref{est8})
one differentiates equation (\ref{pr1})
and uses the estimate 
$\left\Vert\left[F'(x(t))+\ep(t)I\right]^{-1}\right\Vert\le\f{1}{\ep(t)}$. 
The result
is:
\begin{equation}\label{deriv}
||\dot{x}(t)||=|\dot{\ep}(t)|\cdot ||[F'(x(t))+\ep(t)I]^{-1}(x(t)-z_0)||\le 
\f{|\dot\ep(t)|}{\ep(t)}||x(t)-z_0||.
\end{equation}
Here we have used the estimate 
\begin{equation}\label{3.12}
||x(t)-z_0||\leq ||y-z_0||,  
\end{equation}
which can be derived from (\ref{pr1}) similarly to the derivation of (\ref{toobad}).
Thus estimate (\ref{est8}) follows from (\ref{deriv}) and (\ref{3.12}).
$\Box$

\begin{lemma}
Assume that $\ep=\ep(t)>0$, $F$ is twice Fr\'echet differentiable, condition
(\ref{frder}) holds,
and
\begin{equation}\label{N}
||F'(x)||\le N_1, \q ||F''(x)||\le N_2\q  \forall x\in H.
\end{equation}
Then for the operator $\P$ defined by (\ref{rchn}) and $x(t)$, the
solution to (\ref{pr1}) with $\ep=\ep(t)$, estimate (\ref{semon}) holds with
\begin{equation}\label{sigma}
\a(t)\equiv 0,\q \g(t)\equiv 1,\t{ and } \s(t):=\f{N_2}{2\ep(t)}. 
\end{equation}
\label{l4}\end {lemma}  

{\bf Proof} 
Since $x(t)$ is the solution to (\ref{pr1})
applying Taylor's formula one gets: 
$$
(\P(h,t),h-x(t))=-([F'(h)+\ep(t)I]^{-1}[F(h)-F(x(t))+\ep(t)(h-x(t))],h-x(t))
$$
$$
\leq - \Biggl([F'(h)+\ep(t)I]^{-1}[F'(h)(h-x(t))+\ep(t)(h-x(t))],h-x(t)\Biggl)+
\f{N_2||h-x(t)||^3}{2\ep(t)}
$$
\begin{equation}\label{comp}
= -||h-x(t)||^2+\f{N_2||h-x(t)||^3}{2\ep(t)}.
\end{equation}
From (\ref{comp}) and (\ref{semon}) the conclusion of Lemma~\ref{l4}
follows.
$\Box$

Let us state the main result of this section.

\begin{theorem}
Assume:
\begin{enumerate}
\item
problem  (\ref{noneq}) has a unique solution $y$;
\item
$F$ is w-continuous, twice
Fr\'echet differentiable and inequalities (\ref{frder}), (\ref{N}) hold;
\item
 $\ep(t)>0$ is continuously differentiable and monotonically tends to
$0$,
\begin{equation}\label{rhoc}
C_\ep:=\max_{t\in[0,\infty)}\f{\ep(0)|\dot{\ep}(t)|}{\ep^2(t)}<1;
\end{equation}
\item
\begin{equation}\label{rho}
\ep(0)>\f{N_2||z_0-y||}{1-C_\ep}
\max\left\{1,\f{2C_\ep}{1-C_\ep}\right\};
\end{equation} 
\item
$\P$ is defined by (\ref{rchn}). 
\end{enumerate}

Then Cauchy problem (\ref{caupr}) has a unique
solution $z(t)$ for $t\in [0,+\infty)$ and
\begin{equation}\label{lap}
||z(t)-x(t)||\le\f{1-C_\ep}{N_2}\ep(t),\q \lim_{t\to+\infty}||z(t)-y||=0.
\end{equation}
\label{t3}\end{theorem}
  
\begin{remark}
First notice that Theorem~\ref{t3} establishes convergence for any initial
approximation point $z_0$ if $\ep(t)$ is appropriately chosen. To make an 
appropriate choice of $\ep(t)$ one has to choose some function $\ep(t)$ satisfying
condition (\ref{rhoc}). Examples of such functions $\ep(t)$ are given below.
One can observe that condition (\ref{rhoc}) is invariant with
respect to a multiplication $\ep(t)$ by a constant. Therefore one can choose $\ep(t)$
satisfying condition (\ref{rho}) by a
multiplication of the original $\ep(t)$ by a sufficiently large constant.
If $\f{|\dot{\ep}(t)|}{\ep^2(t)}$
is not increasing, then in condition (\ref{rho}) 
$C_\ep:=\max_{t\in[0,\infty)}\f{\ep(0)|\dot{\ep}(t)|}{\ep^2(t)}$ can be replaced by
$C_\ep:=\f{|\dot{\ep}(0)|}{\ep(0)}$.
\label{noincon}\end{remark}

\begin{remark}
In order to get an estimate of the convergence rate for $||x(t)-y||$
one has to make some additional assumptions either 
on $F(x)$ or on the choice of the  
initial approximation $z_0$. 
Without such assumptions one cannot give an estimate of  
the convergence rate. Indeed, as a simple example consider the scalar
equation $F(x):=x^m=0$. Then one gets 
the following algebraic equation for $x(\ep)$: 
\begin{equation}\label{algeq} 
F_\ep(x):=x^m+\ep(x-z_0)=0.
\end{equation}
Assume $m$ is a positive integer and $z_0>0$. It is known
that the solution to this equation is an 
algebraic function which can be represented
by the Puiseux series: 
$x=\sum_{j=1}^\infty c_j\ep^{\f{j}{p}}$ in some neighborhood
of zero. Thus $x=c_1\ep^{\f{1}{p}}(1+O(\ep))$ 
as $\ep\to 0$. Now from (\ref{algeq})
one gets:
$$
c_1^m\ep^{\f{m}{p}}(1+O(\ep))+c_1\ep^{1+\f{1}{p}}(1+O(\ep))=z_0\ep.
$$
Thus $p=m$, $c_1=z_0^{\f{1}{m}}$ and 
$x(\ep)=z_0^{\f{1}{m}}\ep^{\f{1}{m}}(1+O(\ep))$.
For $\ep=0$ one gets the solution $y=0$. Therefore
\begin{equation}\label{algeq1}
|x(\ep)-y|\sim \ep^{\f{1}{m}},\q \ep\to 0,
\end{equation}
where $m$ can be chosen arbitrary large.

Below in Propositions~\ref{prop1} and ~\ref{prop2} some 
sufficient conditions are given that allow one to obtain the estimates for
$||x(t)-y||$.
\label{prem}\end{remark}

{\bf Proof} {\bf of Theorem~\ref{t3}} 
Choosing $\mu(t)=\f{\lambda}{\ep(t)}$, where $\lambda$ is a constant, 
from conditions (\ref{ric1}) and (\ref{xdot}) one gets the following inequalities: 
\begin{equation}\label{rho1}
N_2\le\lambda\left(1-\f{|\dot{\ep}(t)|}{\ep(t)}\right),
\end{equation}
\begin{equation}\label{rho2}
2||z_0-y||\f{|\dot{\ep}(t)|}{\ep^2(t)}\le\f{1}{\lambda}
\left(1-\f{|\dot{\ep}(t)|}{\ep(t)}\right),\q \lambda||z_0-x(0)||<\ep(0).
\end{equation}
Choose
\begin{equation}\label{chlam}
\lambda:=\f{N_2}{1-C_\ep}.
\end{equation}
It follows from  (\ref{rhoc}) that (\ref{rho1}) holds.
From (\ref{rho}) one gets:
\begin{equation}\label{ad1}
\ep(0)>\f{N_2||z_0-y||}{1-C_\ep}.
\end{equation}
On the other hand from (\ref{toobad}) it follows that 
\begin{equation}\label{ad2} 
||z_0-x(0)||\le ||z_0-y||.
\end{equation}
Thus from (\ref{ad1}), (\ref{ad2}) and (\ref{chlam}) one obtains the second
inequality in (\ref{rho2}).
Using (\ref{rho}) once again, one gets:
\begin{equation}\label{ad3} 
\ep(0)\geq \f{2N_2C_\ep||z_0-y||}{(1-C_\ep)^2}.
\end{equation}
This inequality and (\ref{rhoc}) imply: 
\begin{equation}\label{ad4}
\lambda=\f{N_2}{1-C_\ep}\le 
\f{1-C_\ep}{2||z_0-y||\f{|\dot{\ep}(t)|}{\ep^2(t)}}.
\end{equation}
By (\ref{rhoc}) one gets
\begin{equation}\label{ad5}
\f{N_2}{1-C_\ep}\le\f{1-\f{|\dot{\ep}(t)|}{\ep(t)}}{2||z_0-y||
\f{|\dot{\ep}(t)|}{\ep^2(t)}}.
\end{equation}
The first inequality in (\ref{rho2}) is equivalent to (\ref{ad5}) for
$\lambda$  chosen in  (\ref{chlam}). Therefore one gets inequality
(\ref{lap}) 
by applying Theorem~\ref{t1}, 
while the second relation (\ref{lap}) follows from
(\ref{znorms}), inequality (\ref{lap}) and the triangle inequality:
$$
||z(t)-y||\le ||z(t)-x(t)||+||x(t)-y||.
$$  
Theorem~\ref{t3} is proved.
$\Box$

{\bf Examples.}

{\bf 1.} Let $\ep(t)=\ep_0(t_0+t)^{-\nu}$, $\ep_0$, $t_0$ and $\nu$ are
positive constants. Then $C_\ep=\f{\nu}{t_0}$ and 
condition (\ref{rhoc})  is satisfied if $\nu\in (0,1]$ and $t_0>\nu$.

{\bf 2.} If $\ep(t)=\frac {\ep_0}{\log(t_0+t)}$, 
then $C_\ep=\f{1}{t_0\log t_0}$ and
condition (\ref{rhoc}) is satisfied if \break $t_0\log t_0>1$.   

Note that if $\ep(t)=\ep_0e^{-\nu t}$ then condition (\ref{rhoc}) is not
satisfied.

\begin{proposition}
{\it Let the assumptions of Theorem~\ref{t3} hold. Suppose also that the
following inequality holds:
\begin{equation}\label{pr11}
(F(h),h-y)\ge c||h-y||^{1+a},\q a>0.
\end{equation}
Then for the solution $z(t)$ to problem (\ref{caupr}) the following
estimate holds: 
\begin{equation}\label{pr12}  
||z(t)-y||=O\left(\ep^{\f{1}{a}}(t)\right).
\end{equation}
}
\label{prop1}\end{proposition}

{\bf Proof}
Denote $||x(t)-y||:=\varrho(t)$. Since $F(y)=0$, inequality  (\ref{3.6}) implies 
\begin{equation}\label{pr13}
c\varrho^{1+a}(t)+\ep(t)\varrho^2(t)\le \ep(t)||z_0-y||\varrho(t)
\end{equation}
and $\varrho(t) \to 0$ as $t\to +\infty$. This inequality can be reduced to 
\begin{equation}\label{pr14}
c\varrho^{a}(t)+\ep(t)\varrho(t)\le \ep(t)||z_0-y||.
\end{equation}
Thus $\varrho^{a}(t)\le \f{||z_0-y||}{c} \ep(t)$, and 
\begin{equation}\label{pr15}
 ||x(t)-y||\le \left( \f{||z_0-y||}{c}\right)^{\f{1}{a}}\ep^{\f{1}{a}}(t).
\end{equation}
 Combining this estimate with estimate (\ref{lap}) for $||z(t)-x(t)||$ one 
completes the proof.
$\Box$

{\bf Example.}
In the case of a scalar function $f(h)$ and even integer $a>0$ the estimate 
$f(h)(h-y)\ge c|h-y|^{1+a}$ means that  $f(h)=(h-y)^{a}g(h)$, 
where $g(h)\ge c>0$, 
and hence $y$ is a zero of the multiplicity $a$ for $f$.

\begin{proposition}
{\it Let all the assumptions of Theorem~\ref{t3} 
hold and there exists $v\in H$ such that
\begin{equation}\label{pr21}
z_0-y=F'(y)v,\q ||v||<\f{2}{N_2}.
\end{equation}
Then for the solution $z(t)$ to problem (\ref{caupr}) the following  
convergence rate estimate holds:
\begin{equation}\label{pr22}
||z(t)-y||\le \left [\f{1-C_\ep}{N_2}+\f{4||v||}{2-N_2||v||}\right ]\ep(t).
\end{equation}
}
\label{prop2}\end{proposition}
  
{\bf Proof}
From (\ref{pr1}) for an arbitrary $\ep>0$ one gets
$$
F(x)-F(y)+\ep(x-y)=\ep(z_0-y).
$$
Therefore by the Lagrange formula one has:
\begin{equation}\label{pr23}
\left\{\il_0^1(F'(y+s(x-y))ds+\ep I\right\}(x-y)=\ep(z_0-y).
\end{equation}
Introduce the notation
$Q_\ep(x):=\il_0^1(F'(y+s(x-y))ds+\ep I$. From (\ref{pr23}) it follows that
$$
||x-y||=\ep||Q_\ep^{-1}(x)Q_0(y)v||\le \ep||Q_\ep^{-1}(x)(Q_0(y)-Q_\ep(x))v||
+\ep||Q_\ep^{-1}(x)Q_\ep(x)v||.
$$
Since $Q_\ep(x)=Q_0(x)+\ep I$, one obtains
$$
||x-y||\le  \ep||Q_\ep^{-1}(x)(Q_0(y)-Q_0(x))v||+\ep||Q_\ep^{-1}(x)\ep v||
+\ep||v||
$$
\begin{equation}\label{pr24} 
\le\f{N_2}{2}||x-y||\,||v||+2\ep ||v||.
\end{equation}
So, from (\ref{lap}), (\ref{pr21}) and (\ref{pr24}) for $\ep=\ep(t)$ and
correspondingly $x=x(t)$ satisfying the assumptions of Theorem ~\ref{t3} one
concludes that estimate (\ref{pr22}) holds.
$\Box$

Now we describe the simple iteration scheme  for solving nonlinear
equation  (\ref{noneq}). Define:
\begin{equation}\label{4.1}
\P(h,t):=-[F(h)+\ep(t)(h-z_0)],\q \ep(t)>0.
\end{equation}

\begin{lemma}
Assume that $F$ is monotone, 
$\P$ is defined by (\ref{4.1}), 
and $x(t)$ is a solution
to problem (\ref{pr1}) with  $\ep=\ep(t)>0,$ $t\in [0,+\infty)$. Then for the
positive function $\g(t):=\ep(t)$ and for
$\s(t)=\a(t)\equiv 0$  estimate (\ref{semon}) holds.
\label{l9}\end{lemma} 

{\bf Proof}
Since $x(t)$ is a solution to problem (\ref{pr1}), by the monotonicity of $F$ one
has:
$$
(\P(h,t),h-x(t))=-(F(h)-F(x(t),h-x(t))-\ep(t)(h-x(t),h-x(t))
$$
\begin{equation}\label{star1}
\le -\ep(t)||h-x(t)||^2.
\end{equation} 
Lemma~\ref{l9} is proved.
$\Box$

Lemma~\ref{l9} together with  Lemma~\ref{stupid} presented below allow  one to
formulate the convergence result
concerning the simple iteration procedure (see Theorem ~\ref{t5}).

\begin{lemma}
Let $\nu(t)$  be integrable on $[0,+\infty)$.
Suppose  that there exists $T\geq 0$ such that $\nu(t)\in C^1[T,+\infty)$
and
\begin{equation}\label{1b}
\nu(t)>0,\q  -\f{\dot{\nu}(t)}{\nu^2(t)}\leq C,\q \hbox{for}\q t\in [T,+\infty).
\end{equation}
Then
\begin{equation}\label{2b}
\lim_{t\to+\infty}\il^t_0\nu(\tau)d\tau = +\infty.
\end{equation}
\label{stupid}\end{lemma}

{\bf Proof}
One can integrate  (\ref{1b}) 
$$
-\il^t_T \f{\dot{\nu}(\tau)}
{\nu^2(\tau)}d\tau\leq \il^t_T Cd\tau,\q t\in [T,+\infty)
$$
and get
$$
\f{1}{\nu(t)}\leq C(t-T) + \f{1}{\nu(T)}.
$$
Without loss of generality we can assume that $C>0$, and then
$$
\nu(t)\geq\f{1}{ C(t-T) + \f{1}{\nu(T)}}.
$$
Integrating this inequality one gets (\ref{2b}) and completes the proof.            
$\Box$

 Lemmas ~\ref{l1} - \ref{l3} and Lemmas ~\ref{l9} - ~\ref{stupid}  imply the
following result.

\begin{theorem}
Assume that:
\begin{enumerate}
\item
problem  (\ref{noneq}) has a unique solution $y$;
\item
$F$ is w-continuous and monotone;
\item
 $F$ is continuously Fr\'echet differentiable and
\begin{equation}\label{N1}
||F'(x)||\le N_1, \q   \forall x\in H;
\end{equation}
\item 
$\ep(t)>0$ is continuously differentiable and tends to zero monotonically as $t\to
+\infty$, and $\lim_{t\to +\infty}\f{\dot{\ep}(t)}{\ep^2(t)}=0$.
\end{enumerate}
Then, for $\P$ defined by (\ref{4.1}), Cauchy problem (\ref{caupr}) has a unique
solution $z(t)$ all for $t\in [0,+\infty)$ and
$$
\lim_{t\to\infty}||z(t)-y||=0.
$$
\label{t5}\end{theorem}

{\bf Proof}
In order to verify the assumptions of Theorem~\ref{t1} we use estimate (\ref{star1})
to conclude that $\a(t)=\s(t)=0$ and $\g(t)=\ep(t)$ in formula (\ref{semon}). By
(\ref{xdot}) $\b(t)=||\dot{x}(t)||$ because $\a(t)=0$. By (\ref{est8}) 
$$\
\b(t)=||\dot{x}(t)||\le\f{|\dot{\ep}(t)|}{\ep(t)}||y-z_0||.
$$
To apply Theorem~\ref{t1} one has to find 
a function $\mu(t) \in C^1[0,+\infty)$
satisfying (\ref{ric1}) that is
\begin{equation}\label{1c}
\f{|\dot\ep(t)|}{\ep(t)}||y-z_0||\leq \f{1}{2\mu(t)}\left(\ep(t)-\f{
\dot\mu(t)}{\mu(t)}\right),\q \mu(0)||x(0)-z_0||<1.
\end{equation}
Such a function can be chosen as the solution to the differential equation 
\begin{equation}\label{2c}
-\f{\dot{\mu}(t)}{\mu^2(t)}+\f{\ep(t)}{\mu(t)}=\f{A|\dot{\ep}(t)|}{\ep(t)},
\end{equation}
where $A:=2||y-z_0||$. Denote $\rho(t):=\f{1}{\mu(t)}.$ Then
$$
\dot{\rho}(t)+\ep(t)\rho(t)=\f{A|\dot{\ep}(t)|}{\ep(t)}
$$
and
\begin{equation}\label{3c}
\rho(t)=\left[A\il^t_0\f{|\dot{\ep}(s)|}{\ep(s)}e^{\il^s_0\ep(\tau)d\tau}ds
+\f{1}{\mu(0)}\right]
e^{-\il^t_0\ep(\tau)d\tau}.
\end{equation}
Since by Lemma~\ref{stupid} $e^{\il^t_0\ep(\tau)d\tau}\to\infty$ as $t\to\infty$
one can apply L'H\^ospital's rule to obtain from (\ref{3c}) and condition 4 
that
$$
\lim_{t\to +\infty}\rho(t)=\lim_{t\to +\infty}\f{|\dot{\ep}(t)|}{\ep^2(t)}=0.
$$
Therefore $\mu(t)$ tends to $+\infty$. To complete the proof
one can take $\mu(0)$ sufficiently small for the second inequality in
(\ref{1c}) to hold. By Theorem~\ref{t1} one concludes that $||z(t)-x(t)||\to 0$
as $t\to +\infty$ and by Lemma~\ref{l2} that $||x(t)-y||\to 0$ as $t\to +\infty$.
Therefore it follows from the estimate:
$$
||z(t)-y||\le ||z(t)-x(t)||+||x(t)-y||,
$$
that $||z(t)-y||\to 0$ as $t\to +\infty$.
$\Box$

\begin{remark}
From the proof it is clear that one can get the estimate
$||z(t)-x(t)||\le\f{1}{\mu(t)}\to 0$ as $t\to +\infty$, and for the term
$||x(t)-y||$ one can get the rate of convergence if some additional assumptions are
made on $F$ or on $z_0$ (see Propositions~\ref{prop1}, ~\ref{prop2}, and also
Remark~\ref{prem}). 
 \end{remark}

\begin{remark} 
The result obtained in Theorem~\ref{t5} is similar to the result in
\cite{al2}.  The assumptions in \cite{al2} are slightly different. The method of
investigation in \cite{al2} is based on a linear differential inequality which is a
particular case of (\ref{ric4}) with $\s(t)\equiv 0$.  This linear differential
inequality has been used often in the literature by many authors.  
\end{remark}

{\bf Examples.}

{\bf 1.} Let $\ep(t)=\ep_0(1+t)^{-\nu}$, $\ep_0$ and $\nu$ are
positive constants. Then the assumptions of Theorem~\ref{t5}
are satisfied if $\nu\in (0,1)$.
 
{\bf 2.} If $\ep(t)=\f{\ep_0}{\log(1+t)}$, then the assumptions of
Theorem~\ref{t5}  are satisfied.
 
If $\ep(t)=\ep_0e^{-\nu t}$ then condition 4 of Theorem~\ref{t5} is not
satisfied.

\section{Regularized Methods for Non-monotone Operators}
\setcounter{equation}{0}
\setcounter{theorem}{0}
\renewcommand{\thetheorem}{4.\arabic{theorem}}
\renewcommand{\theequation}{4.\arabic{equation}}
\vspace*{-0.5pt}
   
In this section we discuss two approaches to the regularization of the 
Gauss-Newton-type  schemes for nonlinear equations with non-monotone operators.
To describe the first one, assume that $F$ in (\ref{noneq}) is compact and 
Fr\'echet differentiable. Denote:
\begin{equation}\label{T}
T(h):=F^{\pr *}(h)F'(h),\q T_{\ep}(h):=T(h)+\ep I.
\end{equation}
Then $T(h)$ is a nonnegative self-adjoint compact operator. Such an
operator cannot be boundedly invertible if $H$ is infinite-dimensional. One
has:
\begin{equation}\label{ep}
|| T^{-1}_{\ep}(h)||\leq \f{1}{\ep}
\end{equation}
for any $\ep >0$. Define $\P$:
\begin{equation}\label{P1}
\P(h,t):=(P_{\ep(t)}(\xi)-I)(h-z_0)-P_{\ep(t)}(\xi)T^{-1}_{\ep(t)}(h)F^{\pr 
*}(h)F(h),\q
\ep(t)>0.
\end{equation}
Here $\xi\in H$ is a fixed element, which will be chosen 
so that inequality  (\ref{sup}) (see below) holds, and
\begin{equation}\label{proj}
P_{\ep}(\xi):=\il_{\ep}^{N_1^2}dE(s),
\end{equation}
$E(s):=E(s,\xi)$ is the resolution of the 
identity of the self-adjoint
operator $T(\xi)$,
and
\begin{equation}\label{N12}
\sup_{h\in H}||F'(h)||\leq N_1,
\end{equation}
so that $||T(\xi)||\le N_1^2$.

\begin{lemma}
Assume that:
\begin{enumerate}
\item
problem  (\ref{noneq}) has a unique solution $y$;
\item
$\P$ is defined by (\ref{P1}), $F$ is compact, twice
Fr\'echet differentiable and inequalities  (\ref{N}) hold;
\item
there exist $\xi\in H$ and  $ \ep_0>0$ such that 
\begin{equation}\label{sup}
C:=\sup_{\ep\in (0,\ep_0]} ||P_{\ep}(\xi)T^{-1}_{\ep}(\xi)
\{T(y)-T(\xi)\}||<\f{1}{2};
\end{equation}
\item
$0<\ep(t)\leq \ep_0$.
\end{enumerate}
Then there exist positive functions  $\a(t)$, $\g(t)$ and
$\s(t)$, $t\in [0,+\infty)$, such that  estimate (\ref{semon}) holds
with $x(t)$ replaced by $y$.  
\label{l41}\end{lemma}

\begin{remark}
Condition (\ref{sup})  contains a priori information
 about a nonlinear operator $F$.
This condition allows one to get a convergence rate
for ill-posed problem (\ref{noneq}).  It is
always satisfied in a well-posed case (for a
boundedly
invertible operator $T$) if $\xi$ is sufficiently  close to $y$.
However it is not clear yet how restrictive this condition is, and how it is related
to other conditions that one has to use in order to prove the convergence of the
process in ill-posed cases. 
\label{wrem}\end{remark}

{\bf Proof} {\bf of Lemma~\ref{l41}}
Using the polar decomposition $F'(h)=U(F^{\pr *}(h)F'(h))^{\f{1}{2}}$, where $U$
is a partial isometry, one gets  $F^{\pr *}(h)=T^{\f{1}{2}}U^*$, and, since
$||U^*||=1$,
one obtains:
\begin{equation}\label{polar}
||T^{-1}_{\ep(t)}(h)F^{\pr *}(h)||\leq||T^{-1}_{\ep(t)}(h)T^{\f{1}{2}}(h)||\leq
\max_{0\le s<+\infty}\f{\sqrt{s}}{s+\ep(t)}=\f{1}{2\sqrt{\ep(t)}}.
\end{equation}
Using (\ref{P1}) and the relation 
\begin{equation}\label{47a}
F(h)=F(h)-F(y)=F'(h)(h-y)+R(y,h),
\end{equation}
where $||R(y,h)||\le\f{N_2}{2}||h-y||^2$,
one gets
$$
(\P(h,t),h-y)=((P_{\ep(t)}(\xi)-I)(h-z_0),h-y) 
-(P_{\ep(t)}(\xi)T^{-1}_{\ep(t)}(h)T(h)(h-y),h-y)
$$
$$
+\f{N_2}{4\sqrt{\ep(t)}}||h-y||^3\leq -||h-y||^2+||(P_{\ep(t)}(\xi)-I)(y-z_0)||
||h-y||
$$
\begin{equation}\label{4.7}
+\ep(t)(P_{\ep(t)}(\xi)T^{-1}_{\ep(t)}(h)(h-y),h-y)+\f{N_2}{4\sqrt{\ep(t)}}||h-y||^3.
\end{equation}
Also one has the following estimates:
\begin{equation}\label{2ep}
||P_{\ep(t)}(\xi)T^{-1}_{\ep(t)}(\xi)||\le\max_{s\ge\ep(t)}\f{1}{s+\ep(t)}=
\f{1}{2\ep(t)}, 
\end{equation}
and
$$
||T(h)-T(y)||\le ||F^{\pr *}(h)[F'(h)-F'(y)]+[F^{\pr *}(h)-F^{\pr *}(y)]F'(y)||
$$
\begin{equation}\label{TT}
\leq 2N_1N_2||h-y||.
\end{equation}
Thus, using the identity $A^{-1}-B^{-1}=-B^{-1}(A-B)A^{-1}$
and  inequalities (\ref{2ep}), (\ref{TT}), 
one obtains:
$$
\ep(t)(P_{\ep(t)}(\xi)T^{-1}_{\ep(t)}(h)(h-y),h-y)=
\ep(t)(P_{\ep(t)}(\xi)T^{-1}_{\ep(t)}(\xi)(h-y),h-y)
$$
$$
+\ep(t)(P_{\ep(t)}(\xi)
\{T^{-1}_{\ep(t)}(h)-T^{-1}_{\ep(t)}(\xi)\}(h-y),h-y)\leq \f{1}{2}||h-y||^2
$$
$$
+||P_{\ep(t)}(\xi)T^{-1}_{\ep(t)}(\xi)||\,||T(h)-T(y)||\,
||h-y||^2+||P_{\ep(t)}(\xi)T^{-1}_{\ep(t)}(\xi)\{T(y)-T(\xi)\}||\,||h-y||^2
$$
\begin{equation}\label{4.8}
\leq\left(\f{1}{2}+C\right)||h-y||^2+\f{N_1N_2}{\ep(t)}
||h-y||^3.
\end{equation}
Define:
\begin{equation}\label{g}
\g(t)\equiv\g:=\f{1}{2}-C;
\end{equation}
\begin{equation}\label{s}
\sigma(t):=\f{N_1N_2}{\ep(t)}+\f{N_2}{4\sqrt{\ep(t)}};
\end{equation}
\begin{equation}\label{a}
\a(t):=||(P_{\ep(t)}(\xi)-I)(y-z_0)||.
\end{equation}
These functions are positive if (\ref{sup}) holds.
If $\epsilon (t)\to 0$ then $\sigma(t)\to +\infty$ and $\alpha(t)\to 0$ as
$t\to +\infty$.
Comparing (\ref{4.7}) - (\ref{a}) with (\ref{semon}) 
and applying Lemma~\ref{ric} one  
completes the proof.  
$\Box$

\begin{theorem}
Suppose that the assumptions of  Lemma~\ref{l41} are satisfied and:
\begin{enumerate}
\item
$P(y-z_0)=0$, where $P$ is an orthonormal projector onto 
the null-space of $T(\xi)$;
\item
$\ep(t)>0$ is continuously differentiable,  monotonically tends to $0$, and
\begin{equation}\label{c0min}
C_0:=\min_{t\in [0,+\infty)}\left\{\g-\f{|\dot{\ep}(t)|}{\ep(t)}\right\}>0,
\end{equation}
with $\g$ defined in (\ref{g});
\item 
$C_{\a}:=\ep(0)\max_{t\in [0,+\infty)}\left\{\f{\a(t)}{\ep(t)}\right\}
<\infty$; 
\item  $\ep(0)$ is chosen so that
\begin{equation}\label{sbg}
\f{4N_1N_2+N_2\sqrt{\ep(0)}}{2C_0}<\ep(0)\min\left\{\f{C_0}{2C_\a},
\f{1}{||y-z_0||}\right\}.
\end{equation}
\end{enumerate}
Then  Cauchy problem (\ref{caupr}) has a unique solution $z(t)$ for
$t\in
[0,\infty)$
and
\begin{equation}\label{n}
||z(t)-y||\le \f{2C_0}{4N_1N_2+N_2\sqrt{\ep(0)}}\ep(t).
\end{equation}
\label{t41}\end{theorem}

\begin{remark}
If $T(\xi)$ is injective then Condition 1 is satisfied automatically. 
\label{nlsp}\end{remark}

\begin{remark}
From (\ref{a}) one can see that $\a$ depends on $y$ and cannot be known a priori.
However it follows from condition 1 of Theorem~\ref{t41} that $\a(t)\to 0$ as $t\to
+\infty$. Therefore in numerical applications of this scheme one should try different
functions $\ep(t)$ (and different points $z_0$) to satisfy condition 3. Then, since
$C_0$ and $C_\a$ are invariant with respect to multiplication of $\ep(t)$ by
a positive constant, one can choose $\ep(0)$ sufficiently large in order to
satisfy condition 4. Such a choice can be done for an arbitrary $z_0$.
\label{yz0}\end{remark}

{\bf Proof} {\bf of Theorem~\ref{t41}} Since $x(t)\equiv y$ in our case, one gets
$\beta(t)=\a(t)$. Let us choose $\mu(t):=\f{\lambda}{\ep(t)}$. Conditions of
Theorem~\ref{t1} can be written as follows:
\begin{equation}\label{gnc1}
\f{N_1N_2}{\ep(t)}+\f{N_2}{4\sqrt{\ep(t)}}\le
\f{\lambda}{2\ep(t)}\left(\g-\f{|\dot{\ep}(t)|}{\ep(t)}\right),
\end{equation}
\begin{equation}\label{gnc2} 
\a(t)\le\f{\ep(t)}{2\lambda}\left(\g-\f{|\dot{\ep}(t)|}{\ep(t)}\right),\q
\lambda||y-z_0||<\ep(0).
\end{equation}
Inequality (\ref{gnc1}) is equivalent to the following one:
\begin{equation}\label{gnc1a}
\lambda\ge \f{4N_1N_2+N_2\sqrt{\ep(t)}}{2\left(\g-\f{|\dot{\ep}(t)|}{\ep(t)}\right)}.
\end{equation}
Take
\begin{equation}\label{lmd}
\lambda:=\f{4N_1N_2+N_2\sqrt{\ep(0)}}{2C_0}.
\end{equation}
Then (\ref{gnc1a}) follows from (\ref{lmd}), from the monotonicity of $\ep(t)$ and
from (\ref{c0min}).
Inequality  (\ref{sbg})  implies that 
\begin{equation}\label{4.18c}
\f{4N_1N_2+N_2\sqrt{\ep(0)}}{2C_0}<\f{C_0}{2\max_{t\in
[0,+\infty)}\left\{\f{\a(t)}{\ep(t)}\right\}}
\end{equation}
If $\lambda$ is defined by (\ref{lmd}), then one has:
$$
\max_{t\in
[0,+\infty)}\left\{\f{\a(t)}{\ep(t)}\right\}<\f{C_0}{2\lambda}.
$$
From (\ref{c0min}) it follows that the first inequality in (\ref{gnc2}) holds.
Finally, one obtains from (\ref{sbg}) that 
$$
\f{4N_1N_2+N_2\sqrt{\ep(0)}}{2C_0}<\f{\ep(0)}{||y-z_0||},
$$
which implies the second inequality in (\ref{gnc2}).
Since $x(t)\equiv y$ and $\mu(t):=\f{\lambda}{\ep(t)}$, by (\ref{est2})
and
(\ref{lmd}) one concludes that (\ref{n}) holds.    
$\Box$

Consider now a variant of Gauss-Newton continuous
method with $\P$ defined as follows:
\begin{equation}\label{P2}
\P(h,t):=-T^{-1}_{\ep(t)}(h)\{F^{\pr *}(h)F(h)+\ep(t)(h-z_0)\}.
\end{equation}
Here $F$ is not assumed compact. The following lemma is a consequence of
Theorem 2.3 in \cite{ars} (see also Theorem 2.4 in \cite{bns}).

\begin{lemma}
Assume that:
\begin{enumerate}
\item
problem  (\ref{noneq}) has a unique solution $y$;
\item
$\P$ is defined by (\ref{P2}), $F$ is twice
Fr\'echet differentiable and estimates (\ref{N}) hold;
\item
$z_0$ is chosen so that, for some $v\in H$,
one has:  
\begin{equation}\label{istok}
y-z_0=T^\zeta(y)v,\q \f{1}{2}\le\zeta\le 1;
\end{equation}
\item
$\ep(t)>0$,
$$
\left[\f{1}{2}\ep^{\zeta-\f{1}{2}}(t)N_1\zeta^\zeta(1-\zeta)^{1-\zeta}+
\f{N_1^{2(\zeta+1)}}{N_1^2+\ep(t)}\right]||v||<1,
$$
where $$(1-\zeta)^{1-\zeta}|_{\zeta=1}:=\lim_{\zeta\to
1-0}(1-\zeta)^{1-\zeta}=1.$$ 
\end{enumerate}
Then for $x(t)\equiv y$ estimate (\ref{semon}) holds with
\begin{equation}\label{alsig}
\a(t):=\ep^\zeta(t)\zeta^\zeta(1-\zeta)^{1-\zeta}||v||,\q
\s(t):=\f{N_2}{4\sqrt{\ep(t)}},
\end{equation}
\begin{equation}\label{gam}
\g(t):=1-\f{1}{2}\ep^{\zeta-\f{1}{2}}(t)N_2\zeta^\zeta(1-\zeta)^{1-\zeta}||v||
-\f{N_1^{2(\zeta+\f{1}{2})}N_2||v||}{N_1^2+\ep(t)}>0.
\end{equation}
\label{l42}\end{lemma}

\begin{remark}
The convergence theorem in the case $\zeta=1$ for 
continuous Gauss-Newton method is
proved in \cite[Theorem 2.3]{ars}.
For the discrete Gauss-Newton method the case $0\le\zeta <\f{1}{2}$ is analyzed
in \cite{bns} under some additional assumptions on the operator. 
For the                               
noise-free case it is shown that the rate of convergence is $o(\ep^{\zeta})$.
\label{bnsrem}\end{remark}

{\bf Proof} {\bf of Lemma~\ref{l42}}
From (\ref{47a}) and (\ref{istok}) one gets 
$$
(\Phi(h,t),h-y)=-(T_{\ep(t)}^{-1}(h)[F^{\pr *}(h)F'(h)+\ep(t)(h-y)+\ep(t)(y-z_0)],
h-y)
$$
$$
\le -||h-y||^2+\f{N_2}{4\sqrt{\ep(t)}}||h-y||^3-
\ep(t)(T_{\ep(t)}^{-1}(h)T^\zeta(y)v,h-y).
$$
Following \cite{bns} one estimates the inner product: 
$$
(T_{\ep(t)}^{-1}(h)T^\zeta(y)v,h-y)=(T_{\ep(t)}^{-1}(h)[T(y)-T(h)]T_{\ep(t)}^{-1}(y)
T^\zeta(y)v,h-y)+(T_{\ep(t)}^{-1}(y)T^\zeta(y)v,h-y).
$$
From the spectral theorem for selfadjoint
linear operator $T(\eta)$, and for any $\eta\in H$, one gets:
\begin{equation}\label{spectral}
||T_{\ep(t)}^{-1}(\eta)T^\zeta(\eta)||\le\max_{s\ge
0}\f{s^\zeta}{s+\ep(t)}=
\f{\zeta^\zeta(1-\zeta)^{1-\zeta}}{\ep^{1-\zeta}(t)}.
\end{equation} 
Since 
$$
T(h)-T(y)=F^{\pr *}(h)[F'(h)-F'(y)]+[F^{\pr *}(h)-F^{\pr *}(y)]F'(y)
$$
and $\f{1}{2}\leq\zeta\le 1$,  
from the polar decomposition one gets  the estimate
$$
||T^{\f{1}{2}}(\eta)T_{\ep(t)}^{-1}(\eta)T^\zeta(\eta)||
\le\max_{N_1^2\ge s\ge
0}\f{s^{\zeta+\f{1}{2}}}{s+\ep(t)}
\le\f{N_1^{2(\zeta+\f{1}{2})}}{N_1^2+\ep(t)},
$$ 
which implies
$$
||T_{\ep(t)}^{-1}(h)[T(y)-T(h)]T_{\ep(t)}^{-1}(y)T^\zeta(y)||
\le ||T_{\ep(t)}^{-1}(h)F^{\pr *}(h)(F'(h)-F'(y))T_{\ep(t)}^{-1}(y)T^\zeta(y)||
$$
$$
+||T_{\ep(t)}^{-1}(h)(F^{\pr *}(h)-
F^{\pr *}(y))F'(y)T_{\ep(t)}^{-1}(y)T^\zeta(y)||
$$
$$
\le
\left[\f{N_2}{2\sqrt{\ep(t)}}\f{\zeta^\zeta(1-\zeta)^{1-\zeta}}{\ep^{1-\zeta}(t)}+
\f{N_2}{\ep(t)}\f{N_1^{2(\zeta+\f{1}{2})}}{N_1^2+\ep(t)}\right]||h-y||.
$$
Therefore one obtains:
$$
(\Phi(h,t),h-y)\le \ep^\zeta(t)\zeta^\zeta(1-\zeta)^{1-\zeta}||v||\,||h-y||
$$
\begin{equation}\label{es423}
-\left\{1-\f{1}{2}\ep^{\zeta-\f{1}{2}}(t)N_2\zeta^\zeta(1-\zeta)^{1-\zeta}||v||-
\f{N_1^{2(\zeta+\f{1}{2})}N_2||v||}{N_1^2+\ep(t)}\right\}||h-y||^2+\f{N_2}{4\sqrt{\ep(t)}}||h-y||^3.
\end{equation}
$\Box$

\begin{remark}
Note that assumption (\ref{istok}) is not algorithmically verifiable.
However, practitioners may try different $z_0$ and choose the one for which the
algorithm works better, that is convergence is more rapid and the algorithm
is more stable.

Assumptions of the type (\ref{istok}) (sourcewise representation) became popular  
recently, because they allow one to establish some error estimates for the
approximate solution. But one has to remember that the results based on such
assumptions are of
limited value because one has no algorithm for choosing $z_0$ for which   
(\ref{istok}) holds, and $y$ in (\ref{istok}) is unknown.

If $T=T^*$ is compact and the null space
$N(T)=\{0\}$, then the range $R(T)$ is dense in
$H$, so in any neighborhood of $y$ there are points $z_0$ for which (\ref{istok})
holds. On the other hand, since $R(T)$ is not closed in the same neighborhood there
are also points $z_0$ for which (\ref{istok}) fails to hold. This is why the methods
for solving nonlinear ill-posed problems, based on the assumption (\ref{istok}) or
similar assumptions are not quite satisfactory although they might work in
practice sometimes, for reasons which are yet not clear.

In general, in order to get a convergence theorem in an
ill-posed case one needs some additional assumptions on the Fr\'echet derivative of
the operator $F$, for example condition (\ref{sup}), or (\ref{istok}), or
some other condition ( see e.g., \cite{des}, condition (2.11)).
\end{remark}
  
\begin{theorem}
Let the assumptions of  Lemma~\ref{l42} be satisfied and
\begin{enumerate}
\item
 $\ep(t)$  is continuously differentiable, monotonically tends to $0$, and
\begin{equation}\label{C0}
C_0:=\min_{t\in [0,+\infty)}\left\{\g(t)-\zeta\f{|\dot{\ep}(t)|}{\ep(t)}\right\}>0,
\end{equation}
with $\g(t)$ defined in (\ref{gam});
\item $z_0$ and $\ep(0)$ are chosen so that 
\begin{equation}\label{min} 
\f{N_2\ep^{\zeta-\f{1}{2}}(0)}{2C_0}<\min
\left\{\f{C_0}{2\zeta^\zeta(1-\zeta)^{1-\zeta}||v||},
\f{\ep^\zeta(0)}{||y-z_0||}\right\}.
\end{equation}
\end{enumerate}
Then the Cauchy problem (\ref{caupr}) has a unique solution $z(t)$ for
$t\in [0,+\infty)$ and
\begin{equation}\label{nn}
||z(t)-y||\le \f{2C_0}{N_2\ep^{\zeta-\f{1}{2}}(0)}\ep^\zeta(t).
\end{equation}
\label{t42}\end{theorem}

{\bf Proof}
Choose $\mu(t):=\f{\lambda}{\ep^\zeta(t)}$. Conditions of Theorem~\ref{t1} can be
rewritten as follows:
\begin{equation}\label{perv}
\f{N_2}{4\sqrt{\ep(t)}}\le\f{\lambda}{2\ep^\zeta(t)}
\left(\g(t)-\zeta\f{|\dot{\ep}(t)|}{\ep(t)}\right),
\end{equation}
\begin{equation}\label{vtor}     
\zeta^\zeta(1-\zeta)^{1-\zeta}||v||\le\f{1}{2\lambda}
\left(\g(t)-\zeta\f{|\dot{\ep}(t)|}{\ep(t)}\right),\q
\f{\lambda||y-z_0||}{\ep^\zeta(0)}<1.
\end{equation}
Inequality (\ref{perv}) is equivalent to the following one: 
\begin{equation}\label{4.27a}
\f{N_2\ep^{\zeta-\f{1}{2}}(t)}{2\left(\g(t)-\zeta\f{|\dot{\ep}(t)|}
{\ep(t)}\right)}\leq \lambda.
\end{equation}
If one takes 
\begin{equation}\label{4.27b}
\lambda:=\f{N_2\ep^{\zeta-\f{1}{2}}(0)}{2C_0},
\end{equation}
then (\ref{4.27a}) follows from  (\ref{4.27b}), 
 from the monotonicity of $\ep(t)$ and  
from  (\ref{C0}). Inequality  (\ref{min}) implies that 
\begin{equation}\label{4.27c}
\f{N_2\ep^{\zeta-\f{1}{2}}(0)}{2C_0}< 
\f{C_0}{2\zeta^\zeta(1-\zeta)^{1-\zeta}||v||}.
\end{equation}
For $\lambda$ defined by (\ref{4.27b}) 
inequality (\ref{4.27c}) can be written as 
\begin{equation}\label{4.27d}
\zeta^\zeta(1-\zeta)^{1-\zeta}||v||<\f{C_0}{2\lambda}.
\end{equation}
From (\ref{4.27d}) one obtains 
the first inequality  (\ref{vtor}). Finally, from 
(\ref{min}) one concludes that 
$$
\f{N_2\ep^{\zeta-\f{1}{2}}(0)}{2C_0}<
\f{\ep^{\zeta}(0)}{||y-z_0||},
$$
which is equivalent to the second inequality 
 (\ref{vtor}) for  $\lambda$ defined
by (\ref{4.27b}). Since $x(t)\equiv y$ and 
 $\mu(t):=\f{\lambda}{\ep^\zeta(t)}$, by 
 (\ref{est2}) and  (\ref{4.27b}) one gets estimate  (\ref{nn}).  
$\Box$

\begin{remark}
One can take $x(t)$ in Theorem~\ref{t42} as the minimizer of the problem
\begin{equation}\label{argmin}
||\,F(x)||^2+\ep(t)||x-z_0||^2=\inf,\,\, x\in H,\q 0<\ep(t)\to 0,\t{
as }t\to +\infty,
\end{equation}
instead of taking $x(t)\equiv y$. Problem (\ref{argmin}) is solvable for
{\it w}-continuous operator $F$ and $\lim_{t\to\infty}||x(t)-y||=0$. If
one can obtain the estimate: 
\begin{equation}\label{st}
||F(x(t))||=O(\ep(t)),
\end{equation}
then one can prove that
\begin{equation}\label{star}
||\dot{x}(t)||=O\left(\f{|\dot{\ep}(t)|}{\ep(t)}\right)
\end{equation}
and estimate (\ref{semon}) holds. However we do not have examples of
nonlinear
operators $F(x)$ satisfying estimate (\ref{st}) and such that $F'(y)$ is not
boundedly
invertible.
\end{remark}

\section{Numerical Example}
\setcounter{equation}{0}
\setcounter{theorem}{0}
\renewcommand{\thetheorem}{5.\arabic{theorem}}
\renewcommand{\theequation}{5.\arabic{equation}}
\vspace*{-0.5pt}

The aim of this section is to illustrate the efficiency of scheme (\ref{caupr})
with $\P$ defined in (\ref{rchn})  and in  (\ref{P2}) for solving a practically
interesting ill-posed nonlinear equation. 

Consider the following iterative process
$$
x^{k+1}=f_{\mu,z}(x^k)
$$
for a differentiable function $f_{\mu,z}$ depending on two parameters $\mu$ and $z$
with the
only maximum at the point $\hat{x}$. This process is characterized by the Feigenbaum
constants:
$$
\delta_z=\lim_{j\to\infty}\f{\mu_{z_j}-\mu_{z_{j-1}}}{\mu_{z_{j+1}}-\mu_{z_{j}}},\q
\alpha_z=\lim_{j\to\infty}\f{d_{z_j}}{d_{z_{j+1}}},
$$
 where $\mu_j$ are the critical values of parameter $\mu$, for which a doubling of 
the period of the function $f_{\mu,z}$ occurs (the appearance of $2^j$ cycle), and
$d_{z_j}$ is the algebraic
distance (could be negative) between zero and the nearest attractor (the limit point
in $2^j$ cycle).  The calculation of $\a_z$ is a problem of a practical interest
because
it is not known yet if they satisfy any algebraic relations or not. 
As it is shown in \cite{fig}, $\alpha_z$ can be found from the following
nonlinear
functional equation:
\begin{equation}\label{feig}
g_z(x)=-\a_zg_z\left(g_z\left(\f{x}{\a_z}\right)\right)
\end{equation}
with the unknown function $g_z$ and the initial condition $g(0)=1$. Then 
\begin{equation}\label{univ1}
\a_z=-\f{1}{g_z(1)}.
\end{equation}
Functional equation (\ref{feig}) does not have in general a unique solution. In 
\cite{briggs} numerical results are given which suggest, according to \cite{briggs},
that in certain restricted classes of analytic functions the solution to
(\ref{feig}) is unique.

In \cite{briggs} the
constants $\alpha_z$  are computed with high accuracy on a class of even concave
functions 
analytic on $[-1,1]$ for integer $z$, $2\le z\le 12$. Approximate solutions of
(\ref{feig}) are constructed as polynomial approximations:
$$
g_z(x)=1+\sum_{i=1}^nq_i|x|^{zi},
$$
where $q_i$ are the solutions to the following nonlinear system:
\begin{equation}\label{fj}
F_j:=\left(1+\sum_{i=1}^nq_i\right)\left(1+\sum_{i=1}^nq_i|x_j|^{zi}\right)-
1-\sum_{i=1}^nq_i\left|1+\sum_{i=1}^nq_i\biggl|(1+\sum_{i=1}^nq_i)x_j\biggl|^{zi}
\right|^{zi}=0,
\end{equation}
where $x_j$ are obtained from the partition of the segment $[0,1]$.

In \cite{briggs} classical Newton's method is successfully applied to  a numerical
solution of system (\ref{fj})  and computation of $\a_z$ for $z=2,\dots,12$. 
For $z>12$  constants $\a_z$ are not found in  \cite{briggs}.

The goal of our experiment is to calculate  $\a_z$ for  $z=2,\dots,12$ (and to 
compare with \cite{briggs}) and also for $z>12$ using schemes (\ref{caupr})
-- (\ref{rchn})  and (\ref{caupr}) --  (\ref{P2}). The function
$g_z(x)$ is
even,  therefore it is sufficient to find it on
$[0,1]$. Since uniform partition of $[0,1]$ works for small $z$ only ($z=2,3$), the
nonlinear partition $x_j:=\left(\f{j}{n}\right)^{\f{1}{z}}$ is chosen.

The Jacobi matrix
$$
F':=\left[\f{\partial F_j}{\partial q_l}\right]_{j,l=\overline{1,n}}=
|x_j|^{zl}\left(1+\sum_{i=1}^nq_i\right)+\left(1+\sum_{i=1}^nq_i|x_j|^{zi}\right)
-\left|1+\sum_{i=1}^nq_i\biggl|(1+\sum_{i=1}^nq_i)x_j\biggl|^{zi}\right|^{zl}
$$
$$
-\sum_{i=1}^nq_izi\left|1+\sum_{i=1}^nq_i\biggl|(1+\sum_{i=1}^nq_i)x_j\biggl|^{zi}
\right|^{zi-1}\left|\biggl|\biggl(1+\sum_{i=1}^nq_i\biggl)x_j\biggl|^{zl}+
\sum_{i=1}^nq_izi\biggl|\biggl(1+\sum_{i=1}^nq_i\biggl)x_j\biggl|^{zi-1}\right|
$$
is strictly ill-posed for $z\ge 2$ and $n\ge 2$. The condition number for any fixed
$n$ increases about ten times as $z$ is replaced by $z+1$. 
Therefore solving (\ref{fj}) for large $z$ and $n\ge 2$ is a very unstable problem to
which the standard numerical methods are not applicable. However our methods, based on
 Theorems~\ref{t1}, ~\ref{t3} and ~\ref{t42}, do work and yield
the Feigenbaum constants $\a_z$ for $z=2,...,26$.  

For a more accurate 
approximation of 
$g_z$ one has to take $n$ large enough, but then the problem of the choice of an
initial approximation occurs: for $z=2$, $n=2$ or $n=3$ system (\ref{fj}) has many
solutions. By this reason the scheme described  in  \cite{b} is used. First, system
(\ref{fj}) is solved for $n=1$, then the solution of (\ref{fj}) with  $n=1$ is taken
as the initial guess for the case $n=2$, etc. When $z=2$, $n=1$,  $x=1$ system
(\ref{fj}) is reduced to one algebraic equation with respect to $q_1$:
$$
q_1(q_1+1)(q_1^5+3q_1^4+3q_1^3+3q_1^2+2q_1-1)=0
$$
and the two obvious  solutions are $q_1^{(1)}=0$, $q_1^{(2)}=-1$. Since the function
$g_z(x)$ is even and concave,  the initial condition $g_z(0)=1$ implies $g_z(1)<1$,
that is $1+q_1<1$, $q_1<0$. Therefore one has
to find the negative roots of the equation:
$$
q_1^5+3q_1^4+3q_1^3+3q_1^2+2q_1-1=0.
$$
Such roots are: $q_1^{(3)}=-1.8597174...,$ $q_1^{(4)}=-1.4021968...$\,. 
Thus for the system of two equations ($z=2$, $n=2$) the initial data are:
\begin{enumerate}
\item $q_1=-1, \q q_2=0$;
\item $q_1=-1.8597174..., \q q_2=0$;
\item $q_1=-1.4021968..., \q q_2=0$.
\end{enumerate}
In the first two cases the solutions to (\ref{fj}) ($z=2$, $n=2$) are not concave on
$[-1,1]$. In the third case the graph of the polynomial is concave and 
$$
g_2(x)\approx 1-1.5416948x^2+0.1439197x^4.
$$
For  the
system of three equations ($z=2$, $n=3$) the initial data are:
$$
 q_1=-1.5416948, \q  q_2=0.1439197, \q q_3=0.
$$
Then we continue this process.
The maximum dimension we take is $n=12$. If  $n=13$, the discrepancy is not less
than for $n=12$, and after $n=14$  it grows. 

For $z=3$ we begin the computation with one equation ($n=1$) also. As
the initial
approximation $q_1=-1.4021968$ is taken, that is the solution to (\ref{fj}) with
$z=2$,
$n=1$. The dimension increases step by step till the  discrepancy improves.
For $z=3$, $n=1$ the solution to (\ref{fj}) with  $z=3$, $n=1$ is used, etc.
In our experiment $\a_z$ for  $z=2,\dots,26$ are found. For  $z=2,\dots,12$ they 
coincide with $\a_z$ proposed in  \cite{briggs}.
Below the values of $\a_{13} - \a_{26}$ are presented (the values of $\a_2 - \a_{12}$
can be found in \cite{briggs}). As the exact digits the ones that were the same as
the result of  both regularized procedures were taken.  
$$
\a_{13}=-1.22902, \q \a_{14}=-1.21391, \q \a_{15}=-1.20072, \a_{16}=-1.18910,
$$
$$
\a_{17}=-1.17879, \q \a_{18}=-1.16957, \q \a_{19}=-1.1612, \a_{20}=-1.1537,    
$$
$$
\a_{21}=-1.1469, \q \a_{22}=-1.140, \q \a_{23}=-1.134, \a_{24}=-1.129,
$$
$$
\a_{25}=-1.124, \q \a_{26}=-1.12.
$$

Our numerical results indeed demonstrate the efficiency of procedures 
(\ref{caupr}) -- (\ref{rchn})  and (\ref{caupr}) --  (\ref{P2}) and give  
Feigenbaum's  constants for much larger range than in \cite{briggs}, which is of some
practical interest. Contrary to the original conjecture \cite{fig} our numerical
results confirm the conclusion of \cite{briggs}, which says that the Feigenbaum
constants in fact depend on the parameter $z$. 

\section{Appendix}
\setcounter{equation}{0}
\setcounter{theorem}{0}
\renewcommand{\thetheorem}{6.\arabic{theorem}}
\renewcommand{\theequation}{6.\arabic{equation}}
\vspace*{-0.5pt}
 
Here we prove a lemma about nonlinear differential inequalities.
As we have shown in the previous sections, such inequalities are
very useful in applications.
 
\begin{lemma}
Let $u\in C^1[0,+\infty)$, $u(t)\ge 0$ for $t>0$, and
\begin{equation}\label{ap1}
\dot{u}\le -a(t)f(u(t))+b(t),\q \t{ for }\q t>0,\q u(0)=u_0.
\end{equation}
Assume:
 
1) $a(t),b(t)\in C[0,+\infty)$, $a(t)>0$, $b(t)\ge 0$ for $t>0$,
 
2) $\int^{+\infty}a(t)dt=+\infty$, $\f{b(t)}{a(t)}\to 0$ as $t\to +\infty$,
 
3) $f\in C[0,+\infty)$, $f(0)=0$, $f(u)>0$ for $u>0$,
 
4) there exists $c>0$ such that $f(u)\ge c$ for $u\ge 1$.
 
\noindent Under these assumptions (\ref{ap1}) implies
\begin{equation}\label{descon}
u(t)\to 0\q\t{as}\q t\to +\infty.
\end{equation}
\label{apineq}\end{lemma}
\begin{remark}
This Lemma is essentially Lemma 1 from \cite{al1}. We have added condition
4) and changed the proof slightly. Condition 4) is omitted in \cite{al1}.
Without condition 4 the conclusion of Lemma 1 in \cite{al1} is false as we
show by a counterexample at the end of this Appendix.
Condition 4) is equivalent to the condition $f(w)\to 0$ implies $w\to 0$.

Assumptions about smoothness of $a(t)$ and $b(t)$
in \cite{al1} are not
formulated. In Lemma~\ref{apineq} we assume that these functions are 
continuous and $a(t)>0$. The continuity assumption can be relaxed,
but in applications it is not restrictive, since we deal with the
inequality.
\end{remark}
 
{\bf Proof}
By assumptions 1), 2), the new variable $s=s(t):=\il_0^t a(\tau)d\tau$, maps
$t\in [0,+\infty)$ onto $s\in [0,+\infty)$. Write (\ref{ap1}) as
\begin{equation}\label{ap2}
\f{dw}{ds}\le -f(w(s))+\b(s),\q \t{ for }\q s>0,\q w(0)=u_0,
\end{equation}
where $w(s):=u(t(s))$ and $\b(s)=\f{b(t(s))}{a(t(s))}\to 0$ as $s\to +\infty$.
 
The lemma is proved if one proves
\begin{equation}\label{ap3}
w(s)\to 0\q \t{ as }\q s\to +\infty.
\end{equation}
Let $\kappa(s)\in C[0,+\infty)$ be an arbitrary function such that
$\kappa(s)\ge 0$, $\kappa(s)\to 0$ as $s\to +\infty$ and
$\int^{+\infty}\kappa(s)ds=+\infty$.
For example one can take $\kappa(s)=\f{1}{s+1}$.
Define subsets of ${\bf R}_+:=\{s:s\ge 0\}$ as follows:
\begin{equation}\label{ap4}
E:=\{s: s>0, f(w(s))-\b(s)\le\kappa(s)\},\q F:={\bf R}_+\setminus E.
\end{equation}
{\bf Claim:}
\begin{equation}\label{ap5}
\sup E=+\infty.
\end{equation}
We prove (\ref{ap5}) later.
 
Assuming (\ref{ap5}), consider $s_1\in E$,
$(s_1,s_2)\subset F$. Then
\begin{equation}\label{ap6}
-f(w(s))+\b(s)<-\kappa(s)\t{  for  } s_1<s<s_2.
\end{equation}
From (\ref{ap2}) and (\ref{ap6}) one gets
\begin{equation}\label{ap61}
\f{dw}{ds}< -\kappa(s)\t{  for  } s_1<s<s_2.
\end{equation}
Therefore  for $s_1<s<s_2$ one has
\begin{equation}\label{ap7}
w(s)\le w(s_1)-\int_{s_1}^s\kappa(\tau)d\tau\le w(s_1)\t{  for  } s_1<s<s_2,
\quad s\in F.
\end{equation}
Since $s_1\in E$ one has
\begin{equation}\label{ap8}
f(w(s_1))\le \kappa(s_1)+\b(s_1)\to 0\t{ as } s_1\to +\infty, \quad s_1\in E.
\end{equation}
Here we have used the assumptions $\kappa(s)\to 0$ and
$\beta(s)\to 0$ as $ s\to +\infty$.
 
From  (\ref{ap8}) and assumption 4) it follows, that
\begin{equation}\label{ap9}
w(s_1)\to 0\q\t{ as }\q s_1\to +\infty,\q s_1\in E,
\end{equation}
and from (\ref{ap7}) and (\ref{ap9}) it follows that
\begin{equation}\label{ap10}
w(s)\to 0\q\t{ as }\q s\to +\infty,\q s\in F.
\end{equation}
Thus, to prove Lemma it is sufficient to prove (\ref{ap5}).
 
Suppose (\ref{ap5}) is false, that is $\sup E=s_3<+\infty$. Then
\begin{equation}\label{ap11}
f(w(s))-\b(s)>\kappa(s)\q\t{for}\q s>s_3.
\end{equation}
From (\ref{ap2}) and (\ref{ap11}) one gets
\begin{equation}\label{ap12}
\f{dw}{ds}\le -\kappa(s)\q\t{for}\q s>s_3.
\end{equation}
Thus
\begin{equation}\label{ap13}
w(s)\le w(s_3)-\il_{s_3}^s\kappa(\tau)d\tau\to -\infty\q\t{as}\q s\to
+\infty,
\end{equation}
where we have used the assumption $\int^{+\infty}\kappa(s)ds=+\infty$.
This contradicts the assumption $u\ge 0$ and proves  Lemma~\ref{apineq}.
$\Box$
 
The following example (which is a counterexample to Lemma 1 in \cite{al1})
shows that condition 4) of Lemma~\ref{apineq} is essential.
 
Take
\begin{equation}\label{ap14}
f(u)=\left\{\begin{array}{lcc}
u,\q\t{for}\q 0\le u\le 1,\\
\\
e^{1-u},\q\t{for}\q 1\le u<+\infty,
\end{array}\right.\q a(t)\equiv 1,\q b(t)=\f{3}{t+c},
\end{equation}
where $c>e^{-1}$ is an arbitrary constant.
 
One can check immediately that
\begin{equation}\label{ap15}
u(t)=1+log(t+c)
\end{equation}
satisfies inequality (\ref{ap1}).
The choice $c>e^{-1}$ guarantees that $u(t)>0$ for all
$t\ge 0$. Clearly  $u(t)\to +\infty$ as $t\to +\infty$,
so that conclusion (\ref{descon}) of Lemma~\ref{apineq} is false if
condition 4) is omitted.

\section*{Acknowledgments}  
The authors thank Professor Ya. Alber for useful remarks and Professor V. Vasin
for a discussion of the numerical example.

%

\end{document}